\documentstyle[sprocl]{article}

\input BoxedEPS
\SetRokickiEPSFSpecial  
\HideDisplacementBoxes

\def\bea{\begin{eqnarray}}
\def\eea{\end{eqnarray}}

\newcommand{\beq}{\begin{equation}}
\newcommand{\eeq}{\end{equation}}
\newcommand{\be}{\begin{eqnarray}}
\newcommand{\ee}{\end{eqnarray}}
\newcommand{\ben}{\begin{eqnarray*}}
\newcommand{\een}{\end{eqnarray*}}

\newcommand{\wt}{\widetilde}
\def\({\left( }
\def\){\right) }
\def\[{\left[ }
\def\]{\right] }

\newcommand{\order}{{\cal O}}

\newcommand{\dsl}{\raise.15ex\hbox{$/$}\kern-.57em\hbox{$\partial$}}
\newcommand{\Dsl}{\raise.15ex\hbox{$/$}\kern-.65em\hbox{$D$}}
\newcommand{\Tr}{\mbox{Tr}}

\begin{document}

\title{INSTANTON AND MERON PHYSICS IN LATTICE QCD}

\author{JOHN W. NEGELE}

\address{Center for Theoretical Physics, \\Laboratory
        for Nuclear Science, and Department of Physics, \\
        Massachusetts Institute of Technology, Cambridge, 
        Massachusetts 02139 U.S.A.\\E-mail: negele@mitlns.mit.edu}


\maketitle\abstracts{Lattice field theory provides a quantitative tool to study
the role of  nonperturbative semiclassical configurations in QCD.  This talk briefly
reviews our present understanding of the role of instantons in QCD and describes
in detail new developments in the study of merons on the lattice. 
}

\section{Introduction}

As highlighted by this conference, despite a quarter century of effort,
understanding the nonperturbative  origin of confinement in QCD has remained
an elusive challenge.  In this talk, I will describe how combining  lattice field
theory with familiar semiclassical ideas is providing new insight into the role of
instantons and merons in the QCD vacuum and new tools to understand the
physics of confinement.

 One of the great advantages of the path integral formulation of quantum
mechanics and field theory is the possibility of identifying
non-perturbatively the stationary configurations that dominate the action
and thereby identify and understand the essential physics of complex
systems with many degrees of freedom.  Thus, the discovery of instantons
in 1975~\cite{belavin+75} gave rise to great
excitement and optimism that they were the key to understanding QCD. 
Indeed, in contrast to other many body systems in which the quanta
exchanged between interacting fermions can be subsumed into a potential, it
appeared that QCD was fundamentally different, with topological
excitations of the gluon field dominating the physics and being
responsible for a host of novel and important effects including the
$\theta$ vacuum, the axial anomaly, fermion zero modes, the mass of the
$\eta^\prime$, the chiral condensate, and possibly even confinement.  However,
until lattice QCD became sufficiently sophisticated and
sufficient resources could be devoted to the study of instanton  and meron
physics, it had
not been possible to proceed beyond the dilute instanton gas
approximation and a qualitative exploration of merons~\cite{Callan:1977qs}.

Using lattice field theory, our basic strategy is to reverse the
usual analytical process of calculating
stationary configurations and approximately summing the fluctuations around
them. Rather, we  use Monte Carlo sampling of the path integral for QCD
on a lattice to identify typical paths 
contributing to the action and then work
backwards to identify the smooth classical 
solutions about which these paths are
fluctuating.

\section{The Role of Instantons}

Insight into the  role of instantons and their zero modes from lattice QCD has
been reviewed in detail in ref~\cite{Negele:1999ev}, and I will only summarize
the highlights here.
%
%
The instanton content of
gluon configurations can be extracted by cooling~\cite{Berg:1981nw}, 
and the instanton size is
consistent with the instanton liquid model and the 
topological susceptibility agrees
with the Veneziano-Witten formula~\cite{Negele:1999ev}. We obtain 
striking agreement between vacuum
correlation functions, ground state density-density 
correlation functions, and masses
calculated with all gluons and with only instantons~\cite{Chu:1994vi}.  
Zero modes associated with
instantons are clearly evident in the Dirac spectrum, 
and  account for the $\rho, \pi$,
and $\eta^{\prime}$ contributions to vacuum
correlation functions~\cite{Ivanenko:1998nb}.  Finally, we have observed 
directly quark localization at
instantons in uncooled configurations~\cite{Ivanenko:1998nb}.
The physical picture that arises  corresponds closely to
the physical arguments and instanton models 
in which the zero modes
associated with instantons produce localized quark states, 
and quark propagation
proceeds primarily by hopping between these states~\cite{Schafer:1998wv}.
However, of particular relevance to this conference, it is now clear that 
simple
configurations of instantons, such as a dilute gas~\cite{Callan:1977qs} 
or a random
superposition~\cite{Chen:1999ct},  do not confine color charges. It is 
especially
important to emphasize that even  an
ensemble of instantons  with a distribution of large instantons  $\propto
\rho^{-5}$ yields a static potential which approaches a constant at large 
distance
when one is sufficiently careful in calculating the contributions of very large
Wilson loops~\cite{Chen:1999ct}. Although it is possible that 
instanton  configurations that differ essentially from a random superposition
could lead to confinement, it is interesting to discuss here closely related
topological  gluon excitations known as merons.

%
%
%

\section{Merons on the Lattice}

%
%
%

In the space available in these proceedings, I would like to concentrate on
new advances with James V. Steele~\cite{steele}  in studying merons on the lattice.
Merons are solutions to the classical Yang-Mills field equations
with one-half unit of localized topological charge.
Since they are the (3+1)-dimensional nonabelian 
extension of the 't~Hooft--Polyakov monopole
configuration, which has been shown to lead to confinement in 2+1
dimensions~\cite{Polyakov:1977fu}, they  have long been considered a
possible mechanism for confinement~\cite{Callan:1977qs}.

A purely analytic study of merons has proven intractable
for several reasons.
Unlike instantons, no exact solution exists for more than two
merons~\cite{Actor:1979in}, 
gauge fields for isolated merons fall off too slowly ($A\sim1/x$)
to superpose them, and the field strength is singular.
A patched Ansatz configuration that removes the
singularities~\cite{Callan:1977qs} does not satisfy the classical
Yang-Mills equations, preventing even
calculation of 
gaussian fluctuations around 
a meron pair~\cite{Laughton:1980vv}.
Because of these analytic impasses, we have constructed  smooth, finite action
configurations corresponding to a meron pair and their associated fermion 
zero modes on the lattice.
For simplicity, we restrict our discussion to SU(2) color without loss
of generality.


Two known solutions to the classical Yang-Mills
equations in four Euclidean dimensions
are instantons and merons.
Both have topological charge, can be interpreted as tunneling
solutions, and can be written in the general form 
(for the covariant derivative $D_\mu = \partial_\mu - i A_\mu^a
\sigma^a/2$)
\beq
A_\mu^a(x) =  \frac{2\,\eta_{a\mu\nu} \, x^\nu}{x^2} f(x^2)
\ ,
\label{general}
\eeq
with $f(x^2) = x^2/(x^2+\rho^2)$ for an instanton
and $f(x^2)=\frac12$ for a meron.

Conformal symmetry of the classical Yang-Mills action, in
particular under inversion $x_\mu\to x_\mu/x^2$, shows that
in addition to a meron at the origin, there is 
a second meron at infinity. 
These two merons can be mapped to arbitrary positions, which we define
to be the origin and $d_\mu$.
After a gauge transformation, the gauge field for the two merons
takes the simple form~\cite{deAlfaro:1976qz}
\beq
A_\mu^a(x) =  \eta_{a\mu\nu} 
\[ \frac{x^\nu}{x^2}+ \frac{(x-d)^\nu}{(x-d)^2} \] \ .
\label{conform}
\eeq
Similar to instantons~\cite{Schafer:1998wv}, a meron pair can be
expressed in 
singular gauge by performing a large gauge transformation
about the mid-point of the pair, resulting in a gauge field
that falls off faster at large distances ($A\sim x^{-3}$).
A careful treatment of the singularities shows that the
topological charge density is~\cite{deAlfaro:1976qz}
\beq
Q(x) \equiv 
\frac{\Tr}{16\pi^2} 
 \( F_{\mu\nu} \wt F^{\mu\nu} \)
 = \frac12\, \delta^4(x) + \frac12\, \delta^4(x-d) \ ,
\label{topch}
\eeq
yielding total 
topological charge $Q=1$, just like the instanton.

The gauge field Eq.~(\ref{conform}) has infinite action density 
at the singularities $x_\mu=\{0,\, d_\mu\}$.
Hence, a finite action Ansatz has been suggested~\cite{Callan:1977qs}
\beq
A_\mu^a(x) =  \eta_{a\mu\nu} x^\nu \left\{ 
\begin{array}{cll}
\displaystyle
\frac{2}{x^2+r^2} \ ,\qquad & \sqrt{x^2} < r \ ,
\\[3ex]
\displaystyle
\frac{1}{x^2} \ ,\qquad & r < \sqrt{x^2} < R \ ,
\\[3ex]
\displaystyle
\frac{2}{x^2+R^2} \ ,\qquad & R < \sqrt{x^2} \ .
\end{array}
\right.
\label{caps}
\eeq 
Here, the singular meron fields for $\sqrt{x^2}<r$ and $\sqrt{x^2}>R$
are replaced by instanton caps, each containing topological charge 
$\frac12$ to agree with Eq.~(\ref{topch}).
The radii $r$ and $R$ are arbitrary and we
will refer to the three separate regions defined in
Eq.~(\ref{caps}) as regions I, II, and III, respectively.
The action for this configuration is
$
S= \frac{8\pi^2}{g^2} + \frac{3\pi^2}{g^2} \ln \frac{R}{r} \ ,
$
which shows the divergence in the $r\to0$ or $R\to\infty$ limit.
There is no angular dependence in this patching, and so
the conformal symmetry of the meron pair is retained.
Although this patching of instanton caps is continuous,
the derivatives are not, and so the equations of
motion are violated at the
boundaries of the regions in Eq.~(\ref{caps}).

\begin{figure}[t]
\begin{center}
\vspace*{-1cm}
\leavevmode
\BoxedEPSF{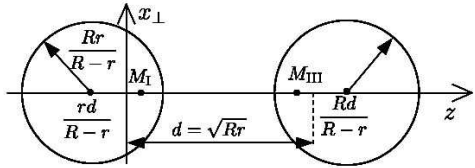 scaled 650}
\end{center}
\caption{\label{fig1} Meron pair separated by $d=\sqrt{R r}$ regulated
with instanton caps, each containing $\frac12$ topological charge.
} 
\vspace{-.4cm}
\end{figure}

Applying the same transformations used to attain Eq.~(\ref{conform})
to the instanton cap solution, regions I and III become
four-dimensional spheres 
each containing half an instanton~\cite{Callan:1977qs}.
The geometry of the instanton caps
is shown in Fig.~\ref{fig1} for the symmetric choice
of displacement $d=\sqrt{R r}$ along the $z$-direction.
Note that the action for the complicated geometry of
Fig.~\ref{fig1} 
with $d=\sqrt{R r}$ can be rewritten as
\beq
S = S_0 \( 1+ \frac{3}{4} \ln \frac{d}{r} \) \ ,
\label{action2}
\eeq
with $S_0 = 8\pi^2/g^2$.
We will concentrate on this case below, and generalization to a different
$d$ is straightforward.

The original positions of the two merons $x_\mu=\{0,\,d_\mu\}$
are not the centers of the spheres, nor are they the positions of
maximum action density, which occurs within the spheres  
with $S_{\rm max} = (48/g^2) (R+r)^4/d^8$ at
\beq
\(M_{\rm I}\)_\mu = \frac{r^2}{r^2+d^2} \; d_\mu \ ,
\qquad
\(M_{\rm III}\)_\mu = \frac{R^2}{R^2+d^2} \; d_\mu \ . 
\label{maxima}
\eeq
However, in the limit $r\to0$ and
$R\to\infty$ holding $d$ fixed, the spheres will shrink around the
original points reducing to the bare meron pair in Eq.~(\ref{conform}). 
In the opposite limit $R\to r$, the radii of the spheres
increase to infinity, leaving an instanton of size $\rho = d$.

An instanton can therefore be interpreted as consisting of a meron
pair. 
This complements the fact that instantons and antiinstantons have
dipole interactions between each other.
If there exists a regime in QCD where meron entropy contributes more
to the free energy than the logarithmic potential between pairs, like
the Kosterlitz--Thouless phase transition, instantons
could break apart into meron
pairs~\cite{Callan:1977qs}. 
The intrinsic size of the instanton caps would be determined by the
original instanton scale $\rho$.  


The Atiyah-Singer index theorem states that 
a fermion in the presence of 
a gauge field with topological charge $Q$, described by the equation
\beq
\Dsl\, \psi = \lambda \psi \ , \quad \mbox{with} \quad
\psi = \( {\psi_R \atop \psi_L} \) \ ,
\label{dirac}
\eeq
has $n_R$ right-handed and $n_L$ left-handed
zero modes (defined by $\lambda=0$) such that $Q=n_L-n_R$.
This statement can often be
strengthened by a vanishing theorem~\cite{Jackiw:1977pu}.
Applying $\Dsl$ twice to $\psi$ decouples the right- and left-hand
components, and focusing on the equation for $\psi_R$, gives
\beq
\( D^2 + \frac12 \bar\eta_{a\mu\nu} \sigma^a F^{\mu\nu} \) \psi_R 
=0\ ,
\label{van}
\eeq
where $\bar\eta_{a\mu\nu}$ denotes the 't~Hooft symbol, $\sigma^a$
acts on the spin indices of $\psi_R$, and $F^{\mu\nu}$ 
acts on the color indices.
For a self-dual gauge field (like an instanton), the second term in
Eq.~(\ref{van}) is zero; and since $D^2$ is a negative definite
operator, there are no 
normalizable right-handed zero modes, implying $Q=n_L$.
Although the meron pair is not self-dual, the second term can be shown
to be negative definite as well, leading to the same conclusion.

In general, a gauge field in Lorentz gauge with $Q=1$ can be
written in the form 
\beq
A_\mu^a(x) = - \eta_{a\mu\nu} \partial_\nu \ln \Pi(x) \ . 
\label{pi}
\eeq
This has a fermion zero mode given by
\beq
\psi = \( {0 \atop \phi} \) \ , \qquad
\phi_\alpha^a = {\cal N} \;  \Pi^{3/2} \;
\varepsilon_\alpha^a \ ,
\label{zmode}
\eeq
with normalization ${\cal N}$ and
$\varepsilon = i \sigma_2$ coupling the spin index $\alpha$ to the
color index $a$ (both of which can be either 1 or 2) in a singlet
configuration.
The gauge field for the meron pair with instanton caps corresponds to
\beq
\!\!\!\Pi(x) = \left\{
\begin{array}{ll}
\displaystyle
\frac{2 \xi_i d^2}{x^2 + \xi_i^2 (x-d)^2}\ , 
& \mbox{for regions $i=$ I, III,}
\\[3ex]
\displaystyle
\frac{d^2}{\sqrt{x^2 (x-d)^2}} \ , 
& \mbox{region II,} 
\end{array}
\right.
\eeq
with $\xi_{\rm I}=r/d$ and $\xi_{\rm III}=R/d$.
The normalized solution to the zero mode is then Eq.~(\ref{zmode})
with 
\beq
{\cal N}^{-1} = 2\pi d^2 \[2-\sqrt{\frac{r}{R}}\]^{1/2}
\ .
\eeq
Note that the unregulated meron pair ($r\to0$, $R\to\infty$)
has a normalizable zero mode itself~\cite{Konishi:1999re} 
\beq
\phi_\alpha^a(x) = \frac{d\; \varepsilon_\alpha^a}
{2\sqrt{2} \pi \( x^2 (x-d)^2 \)^{3/4}} \ .
\eeq
The gauge-invariant zero mode density $\psi^\dagger \psi(x)$ 
has a bridge between two merons that falls off like $x^{-3}$
in contrast to the $x^{-6}$ fall-off in all other directions.
This behavior can be used to identify merons when analyzing their zero
modes on the lattice, similar to what was done for instantons in
Ref.~\cite{Ivanenko:1998nb}. 


\begin{figure}[t]
\begin{center}
\leavevmode
\vspace{-.5cm}
\BoxedEPSF{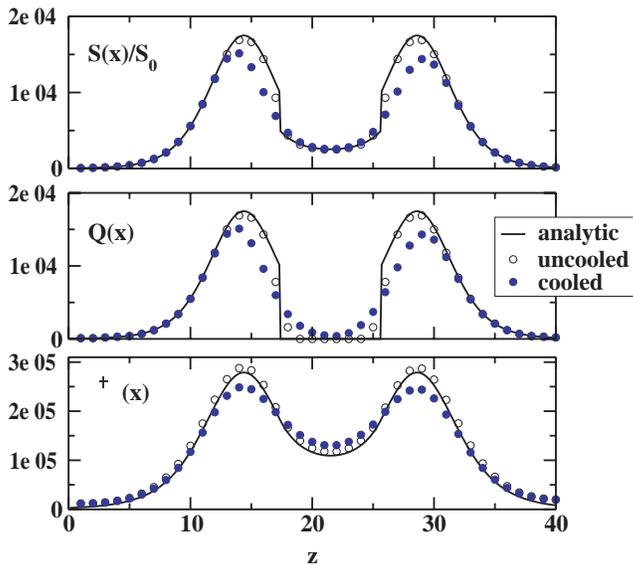 scaled 750}
\end{center}
\caption{\label{fig2} The action density $S(x)$ for the
regulated meron pair,
normalized by the instanton action $S_0=8\pi^2/g^2$, 
sliced through the center of the configuration in the direction of
separation. 
Shown are the analytic (solid), initial lattice (open circles)
and cooled lattice (filled circles) configurations.  The same is shown
for the topological charge density $Q(x)$ and fermion zero mode
density $\psi^\dagger\psi(x)$.}
\vspace{-.35cm}
\end{figure}

As mentioned above, the patching of instanton caps to obtain an
explicit analytic solution has unavoidable and unphysical
discontinuities in the action density.
Therefore, 
in order to study this solution further,
we put the gauge field on a space-time lattice of spacing $a$ in a
box of size $L_0\times L_1\times L_2\times L_3$.
The gauge-field degrees of freedom
are replaced by the usual parallel transport,
$U_\mu(x) = {\rm P} \exp \[-i \int_x^{x+a\, \hat e_\mu} 
A_\nu(z) dz^\nu\]$.
The exponentiated integral can be performed analytically
within the instanton caps, producing arctangents.
Outside of the caps, 
the integral is evaluated numerically by 
dividing each link
into as many sub-links as necessary to reduce the 
$\order(a^3)$ path-ordering errors below machine precision.

\begin{figure}[t]
\begin{center}
\BoxedEPSF{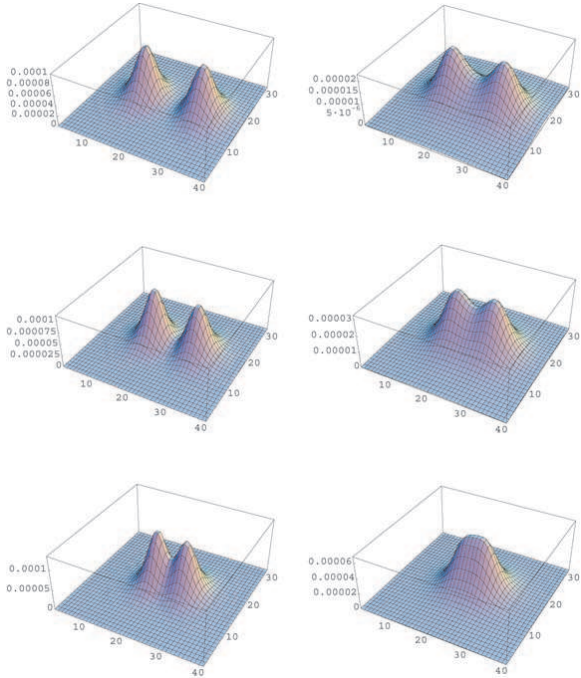 scaled 750}
\end{center}
\caption{\label{fig3} The topological charge density $Q(x)$ on the
left and
fermion zero mode density $\psi^\dagger\psi(x)$ on the right
for a cooled meron pair
with $r=9$, separated by distance $d=21$, 18, and 14 respectively,
in the $(z,t)$-plane. }
\vspace{-.4cm}
\end{figure}

We then calculate the action density $S(x)$ for a meron pair with
instanton caps using the improved action of
Ref.~\cite{GarciaPerez:1994ki} and 
the topological charge density $Q(x)$ using 
products of clovers.
These (open circles) are compared with the patched Ansatz
results (solid line)
in Fig.~\ref{fig2} for $r=9$ and $d=21$ 
(in units of the lattice spacing) on a $32^3\times40$ lattice.
We use the Arnoldi method to solve for the zero mode of this
gauge configuration and hence the density $\psi^\dagger \psi(x)$,
which is also compared with the analytic result in the same figure,
showing excellent agreement.

Two important features of the patched Ansatz also evident with the
lattice representation are the discontinuities in the action density 
at the boundary of the instanton caps and
vanishing of the topological charge density outside of the caps as given
by Eq.~(\ref{topch}).
The discontinuity in the action density is unphysical and can be 
smoothed out by using a
relaxation algorithm to iteratively minimize the lattice
action~\cite{Chu:1994vi}.
On a sweep through the lattice, referred to as a cooling step,
each link is chosen to locally minimize the action density.
Since a single instanton is already a minimum of the action, 
this algorithm would leave the instanton unchanged (for a suitably
improved lattice action).
The result for the regulated meron pair
after ten cooling steps is represented in Fig.~\ref{fig2} by the
filled circles, showing the discontinuities in the action density
have already been smoothed out.
In Fig.~\ref{fig3}, we 
also plot side-by-side the cooled $Q(x)$ and $\psi^\dagger \psi(x)$ 
in the $(z,t)$-plane for three different meron pair separations.
As the separation vanishes, the zero mode
goes over into the well-known instanton zero mode result.

Like instanton-antiinstanton pairs, however, 
a meron pair is not a strict minimum of the action,
since it has a weak attractive interaction Eq.~(\ref{action2})
and under repeated relaxation will coalesce to form an
instanton. 
Nevertheless, just as it is important to sum all the quasi-stationary
instanton-antiinstanton configurations to obtain essential
nonperturbative physics, 
meron pairs may be expected to
play an analogous role.
A precise framework for including quasi-stationary meron pairs is to
introduce a constraint ${\cal Q}[A]$ on a suitable collective variable
$q$ (which we chose to be the quadrupole moment $3z^2-x^2-y^2-t^2$ of
the topological charge)  
as follows
\be
Z &=& \int\!\! DA \; \exp \left\{-S[A]\right\} = \int\!\! dq \; Z_q \ ,
\\
Z_q &=& \int\!\! DA \; \exp \left\{ -S[A] - \lambda \( {\cal Q}[A]-q
\)^2 \right\} \ .
\ee
The meron pair is then a true minimum of the effective action with
constraint, allowing for a semiclassical treatment of $Z_q$.
Afterwards, $q$ is integrated to obtain the full partition function $Z$.

\begin{figure}[t]
\begin{center}
\leavevmode
\BoxedEPSF{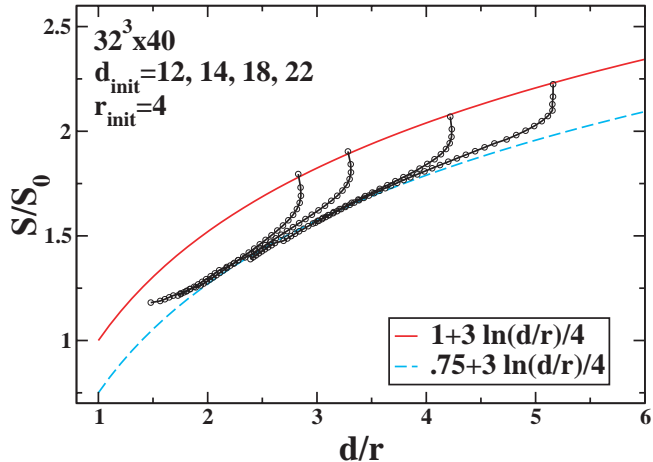 scaled 700}
\end{center}
\caption{\label{fig4} Action of a regulated meron pair 
as a function of $d/r$ for initial lattice configurations of $d_{\rm
init}=12$, 14, 18, 22 and $r_{\rm init}=4$ on a $32^3\times40$ lattice. 
The circles show cooling trajectories for each cooling step.}
\vspace{-.5cm}
\end{figure}
%

The criterion for a good collective variable $q$ of the system is that
the gradient in the direction of $q$ is small compared to the
curvature associated with all the quantum fluctuations.
In this case, we have an adiabatic limit in which 
relaxation of the unconstrained
meron pair slowly evolves through a sequence of
quasi-stationary solutions,
each of which is close to a corresponding stationary constrained solution. 
Detailed comparison of the quasi-stationary and constrained solutions
shows this adiabatic limit is well satisfied.
Therefore, we will present here the action of a
meron pair as it freely coalesces into an instanton.
To compare with the patched Ansatz, we need the meron separation $d$
and radii $r$ and $R$ of the lattice configuration.
We find these by first measuring the separation of the
two maxima in the action density $\Delta\equiv|M_{\rm I}-M_{\rm III}|$ 
(using cubic splines)
and their values (which are the same in the symmetric case,
denoted by $S_{\rm max}=48 /g^2 w^4$),
and then using Eq.~(\ref{maxima}) to give
\beq
d^2 = \Delta^2 + 4 w^2 \ ,
\qquad
\{r,R\} = \frac{2w d}{d\pm\Delta} \ .
\eeq
In Fig.~\ref{fig4}, we plot the total action of a meron pair as a function
of $d/r$ (solid line), which for the analytic case is given by
Eq.~(\ref{action2}). 
Also shown are cooling trajectories for 
four lattice configurations with different
initial meron pair separations.
Each case starts with a patched Ansatz of a given separation and the
lattice action agrees with that of the analytic Ansatz.

The first few cooling steps primarily decrease the action without
changing the collective variable (which is now effectively $d/r$), 
as observed above in Fig.~\ref{fig2}.
Further cooling gradually decreases the collective variable, tracing
out a new
logarithmic curve for the total action 
(dashed line in Fig. 4), which is about 0.25 lower than the analytic case. 
This curve represents the total action of the smooth adiabatic or
constrained lattice meron pair solutions.
The essential property of merons that could allow 
them to dominate the path integral and confine color charge
is the logarithmic interaction Eqs.~(\ref{action2})
which is weak enough to be dominated by the meron 
entropy~\cite{Callan:1977qs}.
Hence, the key physical result from Fig.~\ref{fig4} is the fact that
smooth adiabatic or constrained lattice meron pair solutions clearly
exhibit this logarithmic behavior.

In summary, we have found stationary meron pair solutions, shown that
they have a logarithmic interaction, and shown that they have
characteristic fermion zero modes localized about the individual
merons. 
Hence the presence and role of merons in numerical evaluations of the
QCD path integral should be investigated, and the study of fermionic
zero modes and cooling of gauge configurations provide possible tools
to do so.

\section*{Acknowledgments}
It is a pleasure to thank Professor Hideo Suganuma  for
organizing this conference on confinement and for warm hospitality in
Osaka.  This work was supported in part by the U.S.
Department of Energy under cooperative research agreement
\#DF-FC02-94ER40818. MIT-CTP \#3005

\end{document}